\documentclass[review]{elsarticle}

\usepackage{lineno,hyperref}
\modulolinenumbers[5]

\usepackage[latin1]{inputenc}
\usepackage{natbib}

\usepackage{graphics}
\usepackage{graphicx}

\usepackage[english]{babel}
\usepackage{natbib}
\usepackage{url}

\usepackage{tabularx}
\usepackage{rotating}

\usepackage{xspace}
\usepackage{bm}

\usepackage{amsmath}

\def\RR{\textsf{R}\xspace}

\let\pkg=\strong
\let\proglang=\texttt
\newcommand\code{\bgroup\@codex}
\def\@codex#1{\small {\normalfont\ttfamily\hyphenchar\font=45 #1}\egroup}

\begin{document}

\begin{frontmatter}

\title{Spatial Models with the Integrated Nested Laplace Approximation within Markov Chain Monte Carlo}

\author[uclm]{Virgilio G\'omez-Rubio\corref{cor1}}%\fnref{fn2}}
\ead{Virgilio.Gomez@uclm.es}
\author[uclm]{Francisco Palm\'i-Perales}
\ead{Francisco.Palmi@uclm.es}
\address[uclm]{Department of Mathematics,
School of Industrial Engineering,
University of Castilla-La Mancha,
02071 Albacete, Spain.}

\cortext[cor1]{Corresponding author}
\date{}

\begin{abstract}

The Integrated Nested Laplace Approximation (INLA) is a convenient
way to obtain approximations to the posterior marginals for parameters in
Bayesian hierarchical models when the latent effects can be expressed as a
Gaussian Markov Random Field (GMRF).  In addition, its
implementation in the \pkg{R-INLA} package for the R statistical
software provides an easy way to fit models using INLA in practice.
\pkg{R-INLA} implements a number of widely used latent models, including
several spatial models. In addition, \pkg{R-INLA} can fit models in a fraction
of the time than other computer intensive methods (e.g. Markov Chain Monte
Carlo) take to fit the same model.

Although INLA provides a fast approximation to the marginals of the model
parameters, it is difficult to use it with models not implemented in
\pkg{R-INLA}. It is also difficult to make multivariate posterior inference on
the parameters of the model as INLA focuses on the posterior
marginals and not the joint posterior distribution.

In this paper we describe how to use INLA within the Metropolis-Hastings
algorithm to fit spatial models and estimate the joint posterior distribution
of a reduced number of parameters. We will illustrate the benefits of this new
method with two examples on spatial econometrics and disease mapping where
complex spatial models with several spatial structures need to be fitted.

\end{abstract}


\end{frontmatter}

\section{Introduction}
\label{sec:intro}

Bayesian inference on complex hierarchical models have relied on Markov Chain
Monte Carlo (MCMC, henceforth) methods for many years.  Inference with MCMC is
based on drawing samples from the joint posterior distribution of the model
parameter, and this is often of a high dimension. For Bayesian
hierarchical models with complex structure or large datasets, MCMC can be very
computationally demanding.  See, for example, \citet{Gilksetal:1996} and
\citet{MCMC:2011} for a summary of MCMC methods.

To avoid dealing with high dimensional posterior distributions that are hard to
estimate, \citet{isi:000264374200002} focus on marginal inference and they
develop a method to approximate the posterior marginal of the model parameters
using the Laplace approximation and numerical integration. Also, they focus on
models whose latent effects are a Gaussian Markov Random Field (GMRF,
henceforth).  GMRFs have several properties that can be exploited for
computational efficiency when fitting Bayesian models \citep{RueHeld:2005}.
This new method has been termed the Integrated Nested Laplace Approximation
(INLA, henceforth) and an implementation is available in the \pkg{R-INLA}
package, that can cope with a large family of models.

\citet{Bivandetal:2014} describe a novel approach to fit models not implemented
in \pkg{R-INLA} with INLA.  They note that some models can be fitted with
\pkg{R-INLA} when one of the parameters is fixed, such as several models widely
used in spatial econometrics \citep{LeSagePace:2009}. In this particular case,
the models can be fitted when the spatial autocorrelation parameter is fixed.
As this parameters is in a bounded interval, values for the spatial
autocorrelation parameters can be taken from a fine grid on its (bounded)
support.  Conditioning on these values, models can be fitted to obtain
conditional posterior marginals of all the other parameters. The posterior
marginal of the spatial autocorrelation parameter is then obtained by combining
the marginal likelihoods of the fitted models and the prior using Bayes' rule.
The posterior marginals for the remainder of the parameters in the model is
obtained by averaging over the different conditional posterior marginals.

The former method can be easily extended to more than one parameter but as the
number of parameter increases the number of models to be fitted increases
exponentially. Also, it is problematic when the parameters that need to be fixed
are not in a bounded interval. Hence, if the model needs to be conditioned
on several parameters to be fitted with \pkg{R-INLA}, a different approach
is required.

\citet{GomezRubioRue:2016} suggest using the Metropolis-Hastings algorithm
\citep{Metropolisetal:1953,Hastings:1970} together with INLA when models need
to be conditioned on several parameters. In this way, the joint posterior
distribution of an ensemble of parameters can be obtained via MCMC, whilst the
posterior marginals of all the other parameters is obtained by averaging over
several conditional marginal distributions.

In this paper we extend the work presented in \citet{GomezRubioRue:2016} by
considering specific applications to spatial statistics. In particular, we
consider complex spatial econometrics models and models for disease mapping
with several spatial components.

This paper is structured as follows. Section~\ref{sec:INLA} provides a summary
of INLA and the \pkg{R-INLA} package.  A detailed description of the spatial
models implemented in \pkg{R-INLA} is given in Section~\ref{sec:spINLA}.  Next,
Section~\ref{sec:INLAMCMC} covers how INLA can be used within the
Metropolis-Hastings algorithm to fit new spatial models. This is illustrated in
Section~\ref{sec:examples}, where two examples on spatial econometrics and
joint disease mapping are developed. Finally, a summary and discussion is
available in Section~\ref{sec:discussion}.

\section{Integrated Nested Laplace Approximation}
\label{sec:INLA}

In this Section we will outline the main details about INLA, which is fully
described in \citet{isi:000264374200002}.  Let us assume that we have a vector
$\bm y = (y_1,\ldots,y_n)$ of $n$ observations with a likelihood from the
exponential family.  Mean $\mu_i$ is assumed to be linked, using an appropriate
function,  to a linear predictor $\eta_i$ that will depend on some latent
effects $\bm x$. This is often expressed as

\begin{equation} \eta_i = \alpha + \sum_{j=1}^{n_\beta} \beta_j z_{ji} +
\sum_{k=1}^{n_f} f^{(j)}(u_{ki})+\varepsilon_i
\end{equation}
In this equation, $\alpha$ is an intercept, $\beta_j$ are $n_{\beta}$
coefficients on some covariates $\bm z$, $f^{(k)}$ represent functions on $n_f$
random effects on a vector of covariates $\bm u$, and $\varepsilon_i$ is an
error term.  Hence, the vector of latent effects can be written as 

\begin{equation}
\bm x = \left(\eta_1, \ldots,\eta_n, \alpha, \beta_1, \ldots \right)
\end{equation}
\citet{isi:000264374200002} assume that the structure of the latent effects
$\bm x$ is a Gaussian Markov Random Field \citep[see,][for a detailed
description]{RueHeld:2005}. The precision matrix $\bm Q(\theta_1)$ will
depend on a number of hyperparameters $\theta_1$ and it will fulfill a
number of Markov properties to define the dependences between the elements
in the latent effects. For this reason, $\bm Q(\theta_1)$ is often
very sparse.

Also, observations in $\bm y$ are independent given the values of the latent
effects $\bm x$ and their distribution may depend on a number of hyperparameters
$\theta_2$.

This means that the posterior distribution of the latent effects $\bm x$ and
the vector of hyperparameters $\theta = (\theta_1, \theta_2)$ can be written
down as:

\begin{eqnarray} \pi(\bm x, \theta|\bm y)  = \frac{\pi(\bm y|\bm x,
\theta) \pi(\bm x,\theta)}{\pi(\bm y)}\propto \pi(\bm y|\bm x, \theta) \pi(\bm x, \theta)
\end{eqnarray}
\noindent
$\pi(\bm y)$ is the marginal likelihood and it is a normalizing constant that
is usually ignored because it is often difficult to compute. Nevertheless, 
\pkg{R-INLA} provides an accurate approximation to this quantity that will
play an important role, as we will see later.

In addition, $\pi(\bm y|\bm x, \theta)$ is the likelihood of the model.  As we
are assuming that observations $(y_1,\ldots,y_n)$ are independent given $\bm x$
and $\theta$, then

\begin{equation}
\pi(\bm y|\bm x, \theta) = 
\prod_{i\in \mathit{I}} \pi(y_i |\bm x, \theta)
\end{equation}
$\mathit{I}$ is a set of indices from 1 to $n$ that represents the actually
observed values of $y_i$, i.e., if $y_j$ is missing then $j$ is not
in $\mathit{I}$.

Joint distribution $\pi(\bm x, \theta)$ can be expressed as 
$\pi(\bm x|\theta) \pi(\theta)$. Given that $\bm x$ is a GMRF, we have
the following:

\begin{equation}
\pi(\bm x|\theta) \propto |\bm Q(\theta)|^{1/2}\exp\{-\frac{1}{2}\bm x^T \bm Q(\theta)\bm x\}
\end{equation}

Finally, $\pi(\theta)$ is the prior distribution of the ensemble of
hyperparameters $\theta$. Provided that most of them are independent a priori,
$\pi(\theta)$ can be decomposed as the product of several (univariate)
distributions.

Hence, the joint posterior distributions of the latent effects and parameters
can be written down as

\begin{eqnarray}
\pi(\bm x, \theta|\bm y) \propto  \pi(\theta) |\bm Q(\theta)|^{1/2}\exp\{-\frac{1}{2}\bm x^T \bm Q(\theta)\bm x\}\prod_{i\in \mathit{I}} \pi(y_i |\bm x, \theta)=\nonumber\\
 \pi(\theta) |\bm Q(\theta)|^{1/2}\exp\{-\frac{1}{2}\bm x^T \bm Q(\theta)\bm x + \sum_{i\in \mathit{I}} \log(\pi(y_i |\bm x, \theta))\}
\end{eqnarray}
\indent
It should be noted that because $\bm x$ is a GMRF its precision matrix
$\bm Q(\theta)$ is likely to be very sparse and this can be exploited for
computational purposes.

The posterior marginal distributions of the latent effects can be written as

\begin{equation}
\pi(\bm x_i| \bm y) = \int \pi(\bm x_i, \theta| \bm y) \pi(\theta|\bm y) d\theta
\label{eq:pixi}
\end{equation}
\noindent
Similarly, the posterior marginal of hyperparameter $\theta_i$ can be written as

\begin{equation}
\pi(\theta_i | \bm y) = \int \pi(\theta| \bm y) d\theta_{-i}
\end{equation}
\noindent
where $\theta_{-i}$ is the vector of parameters $\theta$ excluding $\theta_i$.

\citet{isi:000264374200002} propose  an approximation to the joint posterior of
$\theta$, $\tilde\pi(\theta|\bm y)$, that can be used to compute the marginals
of latent effects and parameters:

\begin{equation}
\tilde\pi(\theta|\bm y) \propto \frac{\pi(\bm x,\theta, \bm y)}{\tilde\pi_G(\bm x|\theta, \bm y)} \Big|_{\bm x = \bm x^{*}(\theta)}
\end{equation}
\noindent
Here, $\tilde\pi_G(\bm x|\theta, \bm y)$ is an approximation to the full conditional of $\bm x$ using a Gaussian distribution, and $\bm x^{*}(\theta)$ is
the mode of the full conditional for a given $\theta$.

Approximation $\tilde\pi(\theta|\bm y)$ can be used to  compute
$\pi(\theta_i|\bm y)$ (by integrating $\theta_{-i}$ out) and $\pi(x_i|\bm y)$,
using numerical integration on equation~(\ref{eq:pixi}), but this also requires
a good approximation to $\pi(x_i, \theta| \bm y)$. A Gaussian approximation can
be used as well, but \citet{isi:000264374200002} develop better approximations
using other methods such as the Laplace approximation. 

More details can be found in \citet{BlangiardoCameletti:2015} (Chapter 4),
where computational details are clearly explained and a full example for a
Normal-Gamma model is developed.

INLA is implemented as an \proglang{R} \citep{R:2016} package named
\pkg{R-INLA}.  In addition to providing a simple way of defining models and
computing the approximation to the posterior marginals, \pkg{R-INLA} provides
some extra features for model specification\citep[see,][for
details]{Martinsetal:2013} and can compute a number of derived quantities for
model checking and model assessment . In particular, an approximation to the
marginal likelihood of the model $\pi(\bm y)$ is provided, which can also be
used for model selection, for example.

\section{Spatial Models in INLA}
\label{sec:spINLA}

Several authors
\citep{gomez-rubioetal14,Bivandetal:2015,LindgrenRue:2015,BlangiardoCameletti:2015}
summarize the different spatial models available in \pkg{R-INLA} as latent
effects that can be used to build models.  We will provide a summary in this
Section in order to provide an overview of what has already been implemented
and what other spatial models have not been added yet.

Spatial latent effects for lattice data in \pkg{R-INLA} have a prior
distribution which is a multivariate Normal distribution with zero mean and
precision matrix $\tau T$, where $\tau$ is a precision parameter and $T$ is a
square and symmetric matrix.  The structure of $T$ will control how the spatial
dependence is and it can take different forms to induce different types of
spatial interaction. 

For a completely specified $T$, the multivariate Normal with zero mean and
generic precision matrix $\tau T$ is implemented as the
\texttt{generic0} latent effect.  This can be used to define flexible spatial
structures in case they are not implemented, but matrix $T$ must be fully
defined and it cannot depend on further parameters.

\citet{Besagetal:1991} proposed the use of an intrinsic CAR specification
as a prior for spatial effects that is widely used nowadays. This is
implemented in model \code{besag} and corresponds to the 
precision matrix $\tau Q$, with $Q$ defined as

\begin{equation}
Q_{ij} = 
\left\{
\begin{array}{cc}
n_i & i = j\\
-1 & i \sim j\\
0 & \textrm{otherwise}
\end{array}
\right.
1 \leq i,j \leq n
\end{equation}
\noindent
Here, $n_i$ is the number of neighbors of region $i$, and $i\sim j$ indicates
that regions $i$ and $j$ are neighbors. In this case, spatial random effects
have the additional constrain to sum up to zero in order to make the model
identifiable.

Similarly, the \texttt{generic1} model implements a multivariate Normal with
zero mean and precision matrix $\tau T$, with 

\begin{equation} T = (I - \frac{\beta}{\lambda_{max}}C)
\end{equation}
\noindent
Here, $\beta$ is a parameter which can take values between 0 and 1 and
$\lambda_{max}$ is the maximum eigenvalue of matrix $C$. This latent effect can
be used to implement the model proposed in \citet{Lerouxetal:1999}.  They
propose a model in which the precision matrix is a convex combination of a
diagonal matrix $I$ and matrix $Q$, i.e., the precision matrix is $(1-\lambda)
I  + \lambda Q$ with $\lambda \in (0,1)$.  \citet{Ugarteetal:2014} show that
when $C = I - Q$ then $\lambda_{max} = 1$.  Hence, the model by
\citet{Lerouxetal:1999} can be implemented in \pkg{R-INLA} with a
\code{generic1} model by taking $C = I - Q$, so that $T= I - \beta (I - Q) =
(1-\beta) I + \beta Q$ with $\beta \in (0,1)$.

In addition to the intrinsic CAR specification implemented in model
\code{besag}, model \texttt{bym} implements the sum of an intrinsic CAR and
independent random effects  described in \citet{Besagetal:1991}. Latent model
\texttt{properbesag} implements a proper version of the intrinsic CAR
specification by using a non-singular precision matrix with structure defined
as

\begin{equation}
Q^{\prime}_{ij} = 
\left\{
\begin{array}{cc}
n_i + d & i = j\\
-1 & i \sim j\\
0 & \textrm{otherwise}
\end{array}
\right.
1 \leq i,j \leq n
\end{equation}
\noindent
Here $d > 0$ is an extra parameter that is added to the entries in the diagonal
to make the precision matrix non-singular. Note that this precision matrix
structure $Q^{\prime}$ is the same as in the \texttt{besag} model with an added $d$ to
all the elements in the diagonal so that the resulting distribution is proper,
i.e., $Q^{\prime} = Q +  d I$.

Other models that can be used for spatial modeling are the two dimensional
random walk (named \texttt{rw2d}) and a Gaussian field defined in a regular
lattice with Mat\'ern correlation function (model \texttt{matern2d}). 

All the previous latent models were defined in a lattice, but the \texttt{spde}
model \citep{Lindgrenetal:2011} can be used to define a continuous process with
a Mat\'ern covariance.  This means that it can be used for models in
geostatistics and point patterns \citep{Simpsonetal:2016,GomezRubioetal:2015}.

\citet{Blangiardoetal:2013} and \citet{BlangiardoCameletti:2015} describe 
how to combine these latent effects for spatio-temporal modeling.  In
addition, \citet{Bivandetal:2015} describe how to use the \pkg{INLABMA} package
\citep{Bivandetal:2015} to fit other spatial models using the ideas in
\citet{Bivandetal:2014}. In particular, they fit the model proposed by
\citet{Lerouxetal:1999} and the spatial lag model
\citep{Anselin:1988,LeSagePace:2009}, commonly used in spatial econometrics.  A
new latent effect named \texttt{slm} \citep{GomezRubioetal-slm:2016} is
available (but still experimental) to fit spatial econometrics models.

\section{INLA within Markov Chain Monte Carlo}
\label{sec:INLAMCMC}

As described in the previous Section, \pkg{R-INLA} provides a large number of
spatial latent effects that can be used to build more complex models.  However,
this is list is far from exhaustive and it is difficult to implement new
models in \pkg{R-INLA}. A possible approach to implement new models is the
\texttt{rgeneric} latent effect that allows the user to define the latent model
in \proglang{R} but this requires the user to specify the full structure of the
latent effect as a GMRF.

In order to extend the number of models that INLA can fit through the
\pkg{R-INLA} package, \citet{GomezRubioRue:2016} propose the use of the
Metropolis-Hastings algorithm \citep{Metropolisetal:1953,Hastings:1970}
together with INLA to fit some complex models not implemented in \pkg{R-INLA}.

Let us denote by $\theta$ the ensemble of parameters and hyperparameters to be
estimated in a Bayesian hierarchical models. Note that now $\theta$ will
include some of the latent effects in $\bm x$ and not only the hyperparameters
$\theta_1$ and $\theta_2$ described in Section~\ref{sec:INLA}. Similarly as
\citet{Bivandetal:2014},  \citet{GomezRubioRue:2016} argue that very complex
spatial models can be fitted with \pkg{R-INLA} when some of the parameters are
fixed. Let us call these parameters $\theta_c$ so that $\theta$ becomes
$(\theta_c, \theta_{-c})$.  By conditioning on $\theta_c$, INLA can be used to
obtain the posterior marginals of all the parameters in $\theta_{-c}$, i.e.,
$\pi(\theta_{-c,i}|\bm y, \theta_c)$, and the conditional marginal likelihood
$\pi(\bm y| \theta_c)$.

\citet{Lietal:2012} deal with complex spatio-temporal models in a similar way
by conditioning on some of the parameters at their maximum likelihood
estimates.  Although this is a reasonable way to proceed for highly
parameterized models, it ignores the uncertainty of some of the parameters in
the model when computing the posterior marginal of the other model parameters.
\citet{Bivandetal:2014} also fit spatial models conditioning on some of the
model parameters at different values in a grid and then combine the resulting
models to obtain the posterior marginals (not conditioned on $\theta_c$) for
the model of interest.

The Metropolis-Hastings algorithm could be used to draw values for $\theta_c$
so that its joint posterior distribution can be obtained. At step $i+1$ of the
Metropolis-Hastings algorithm, new values $\theta_c^{(i+1)}$ for $\theta_c$ are
proposed, but not necessarily accepted, from a proposal distribution
$q(\theta_c^{(i+1)}|\theta_c^{(i)})$. The new values are accepted with
probability

\begin{equation}
\alpha =
% \min\{1, \frac{\pi(\theta_c^{(i+1)}|\bm y) q(\theta_c^{(i)}|\theta_c^{(i+1)})}{\pi(\theta_c^{(i)}|\bm y) q(\theta_c^{(i+1)}|\theta_c^{(i)})}\} = 
\min\{1, \frac{\pi(\bm y| \theta_c^{(i+1)}) \pi(\theta_c^{(i+1)}) q(\theta_c^{(i)}|\theta_c^{(i+1)})}{\pi(\bm y|\theta_c^{(i)}) \pi(\theta_c^{(i)}) q(\theta_c^{(i+1)}|\theta_c^{(i)})}\} 
\end{equation}
\noindent
If the proposed value is not accepted, then $\theta_c^{(i+1))}$ is set to
$\theta_c^{(i))}$.

Note that $\pi(\bm y| \theta_c^{(i+1)})$ is the marginal likelihood of the
model conditioned on $\theta_c^{(i+1)}$, which can be obtained by fitting the
model by fixing the values of $\theta_c$ to $\theta_c^{(i+1)}$. Also, this
will provide the posterior marginals $\pi(\theta_{-c,i}|\bm y,
\theta_c^{(i+1)})$ for all the parameters in $\theta_{-c}$.

After a suitable number of iterations, the Metropolis-Hastings algorithm will
produce samples from $\pi(\theta_c|\bm y)$, from which the
the posterior marginals of the parameters in $\theta_c$ can be derived. Also,
we will obtain a family of conditional marginal distributions for all
the parameters in $\theta_{-c}$. Their posterior marginals can be obtained
by combining all these marginals as follows:

\begin{equation}
\pi(\theta_{-c,i}|\bm y) = \int \pi(\theta_{c,i}|\bm y, \theta_c) \pi(\theta_c|\bm y) d\theta_c \simeq \frac{1}{N} \sum_{j = 1}^N \pi(\theta_{c,i}|\bm y, \theta_c^{(j)})
\end{equation}
\noindent
Values $\{\theta_c^{(j)}\}_{j=1}^{N}$ represent $N$ samples from 
$\pi(\theta_c|\bm y)$ obtained with the Metropolis-Hastings algorithm.

\citet{GomezRubioRue:2016} illustrate the use of INLA within the
Metropolis-Hastings algorithm with three examples, including fitting a simple
spatial econometrics model. In addition to fitting models with unimplemented
latent effects, this approach can be employed to use other priors not
implemented in \pkg{R-INLA} (in particular, multivariate priors) and to 
fit models with missing values in the
covariates.  \citet{Camelettietal:2016} discuss how models with missing data
and multiple imputation can be tackled with INLA.

\section{Examples}
\label{sec:examples}

In this Section we provide two examples on how to fit complex spatial models
with INLA using the \pkg{R-INLA} package and the Metropolis-Hastings algorithm.
Both models have a complex spatial structure, with several spatial terms, and
they cannot be fitted with \pkg{R-INLA} alone. In addition, to fitting the
model, we show how to exploit the joint posterior distribution obtained on a
reduced ensemble of parameters to make multivariate inference.

\subsection{Spatial Econometrics} \label{subsec:speco} \citet{Manski:1993}
proposed the following general model for spatial econometrics that accounts for
different levels of spatial dependence and interaction:

\begin{equation}
y = \rho W y + X\beta + W X \gamma + u ;\ u = \lambda W u +e
\end{equation}
$y$ is the response variable made of observations in $n$ areas, $W$ an
adjacency matrix, $X$ a matrix of $p$ covariates, $WX$ a matrix of lagged
covariates, $\beta$ and $\gamma$ associated coefficients, and $\rho$ and
$\lambda$ are spatial autocorrelation parameters.  $e$ is a multivariate
Gaussian error term with
zero mean and diagonal variance-covariance matrix $\sigma^2 I$, where $I$ is
the identity matrix of appropriate dimension.

This model has two autocorrelation terms, one on the response $y$ and another
one on the error term $u$, that are controlled by autocorrelation parameters
$\rho$ and $\lambda$, respectively.  Spatial adjacency matrix $W$ is often
taken as a binary matrix, so that entry at row $i$ and column $j$ will be 1 if
areas $i$ and $j$ are neighbors. Although this is a reasonable adjacency
matrix, we will take $W$ to be row-standardized, so that the sum of all the
elements in a row add up to 1. This will bring interesting properties to the
spatial autocorrelation parameters. In particular, they will be bounded in the
interval $(1/\lambda_{min}, 1)$, where $\lambda_{min}$ is the minimum
eigenvalue of $W$ \citep{Haining:2003}.

\citet{LeSagePace:2009} fit this model using Bayesian inference, for which they
assign priors to all model parameters. In particular, they consider typical
independent inverted gamma priors for all the precisions, vague independent
Normal distributions for $\beta$ and $\gamma$ and uniform prior distribution
for $\rho$ and $\lambda$ in the interval $(1/\lambda_{min}, 1)$.

In order to assess whether Manki's model can be fitted using standard
software for linear regression, it can be rewritten as follows:

\begin{equation}
y = (I - \rho W)^{-1} (X\beta + W X \gamma) + u ; u \sim MVN(0, \Sigma)
\end{equation}
Here, the error term has a multivariate normal distribution with zero mean
and variance-covariance matrix $\Sigma = \sigma^2[(I - \rho W^{\prime}) (I - \lambda W^{\prime}) (I - \lambda W) (I - \rho W) ]^{-1}$.

Hence, this model has a non-linear term on $\rho$, $(I - \rho W)^{-1}$, that
multiplies the linear predictor on the covariates and lagged covariates, and a
complex error structure. This model cannot be fitted with  \pkg{R-INLA} unless
we condition on $\rho$ and $\lambda$.

In order to fit this model with \pkg{R-INLA}, we have combined INLA and MCMC as
described in this paper. In particular, we will be drawing samples from the
posterior of $\theta_c = (\rho, \lambda)$ using the Metropolis-Hastings
algorithm. At each step of the algorithm we will obtain a conditional (on
$\theta_c$) posterior marginal for the remainder of the parameters in the model
$\theta_{-c}$. These conditional marginals will be later combined to obtain the
posterior marginals of the model parameters $\theta_{-c}$.

The proposal distribution for $\theta_c$ has been the product of two Normal
distribution centered at the current value of the parameters and standard
deviation 0.25. This value has been obtained after some tuning to ensure a
suitable acceptance rate. The starting point is $\theta_c = (0,0)$.
%has been the values of $\rho$ and
%$\lambda$ obtained by maximizing the likelihood of the model using function
%\code{sacsarlm()} from package \pkg{spdep}.
Then 500 iterations were used as burn-in, followed by 5000 iterations more, of
which one in 5 was kept to reduce autocorrelation.  This provided a final 1000
samples to estimate the joint posterior distribution of $\theta_c$. Note that
thinning also involves the conditional distributions computed at each step of
the Metropolis-Hastings algorithm.

We have fit this model to the Columbus dataset \citep{Anselin:1988}
which is available in \RR package \pkg{spdep}. This dataset
contains information about 49 neighborhoods in Columbus (Ohio) in 1980.
We will reproduce a classical analysis consisting on explaining crime rate
on housing value (variable HOVAL) and household income (variable INC). We have
not include lagged covariates, so that $\gamma = 0$.

In addition to using INLA within MCMC as described in this paper, we have also
fitted this model using Gibbs sampling 
using the \pkg{rjags} package \citep{rjags:2016} and using maximum
likelihood with function \code{sacsarlm()} from the \pkg{spdep} package
\citep{BivandPiras:2015}.

\begin{figure}
\centering
\includegraphics[width=12cm]{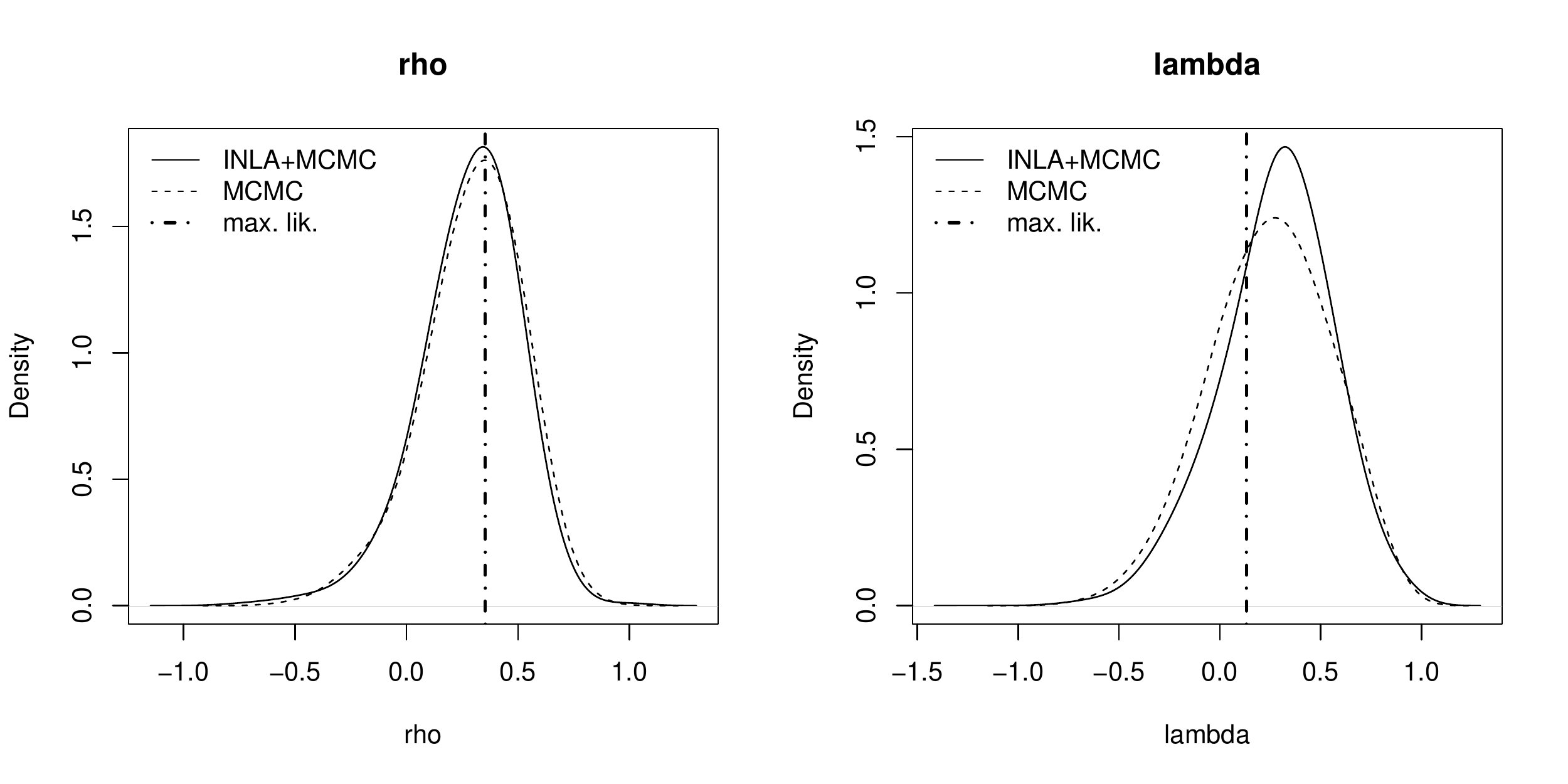}
\caption{Estimates of the spatial autocorrelation parameters
of the Manski model using INLA within MCMC, MCMC and maximum likelihood.}
\label{fig:rholambda}
\end{figure}

\begin{figure}
\centering
\includegraphics[width=12cm]{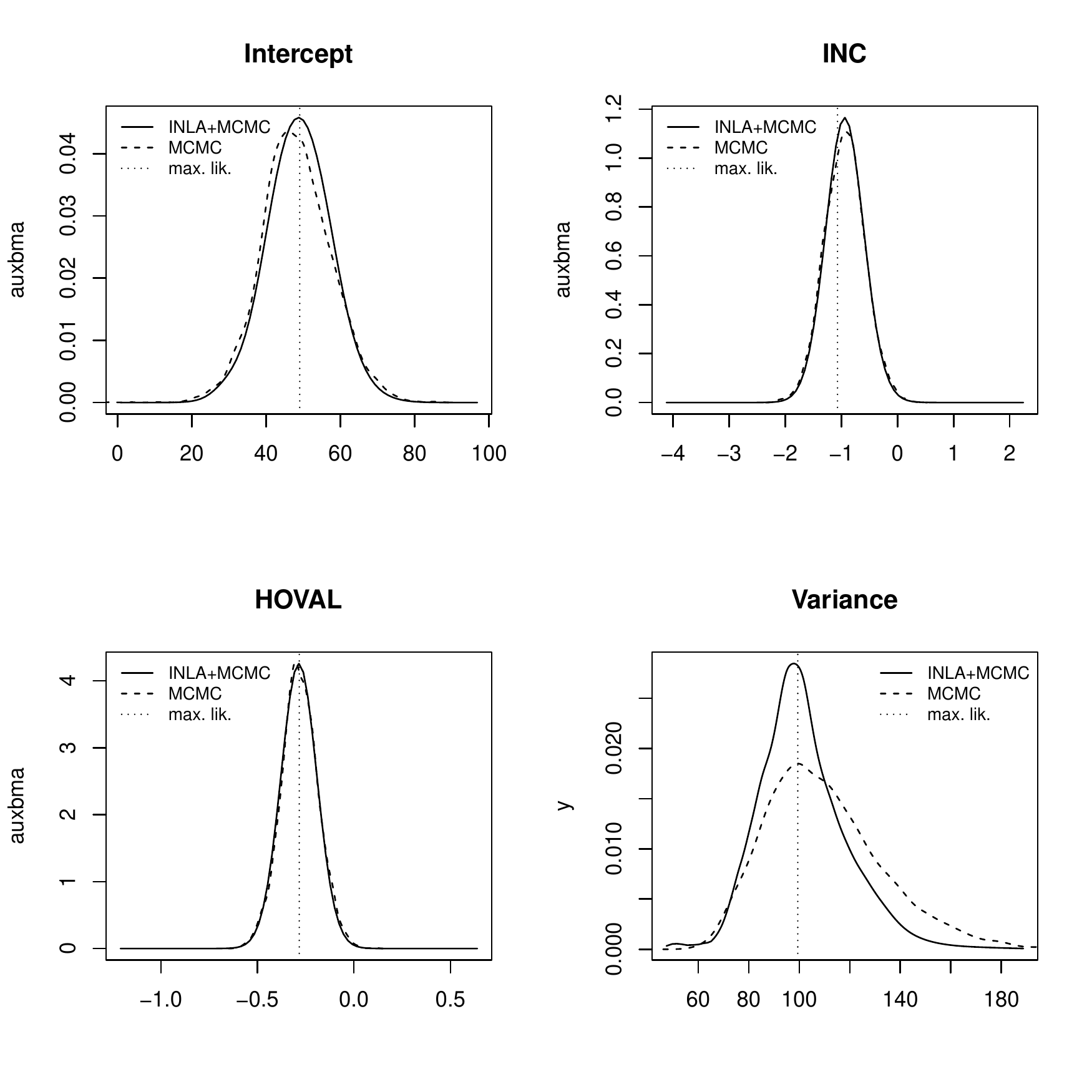}
\caption{Estimates of the intercept, covariate coefficients and variance
using INLA within MCMC, MCMC and maximum likelihood. Covariates are
housing value (HOVAL) and household income (INC).}
\label{fig:betatau}
\end{figure}

Figure~\ref{fig:rholambda} shows the posterior marginals distributions of the
autocorrelation parameters, estimated using MCMC and INLA within MCMC, and 
the maximum likelihood estimates of the parameters. As it can be seen, there is
good agreement between the different estimates.  Regarding the other
parameters in the model, Figure~\ref{fig:betatau} summarizes the different
estimates and how they seem to agree. Note that
these marginals have been obtained by averaging the conditional marginals
obtained at each step of the Metropolis-Hastings algorithm.

\begin{figure}
\centering
\includegraphics[width=12cm]{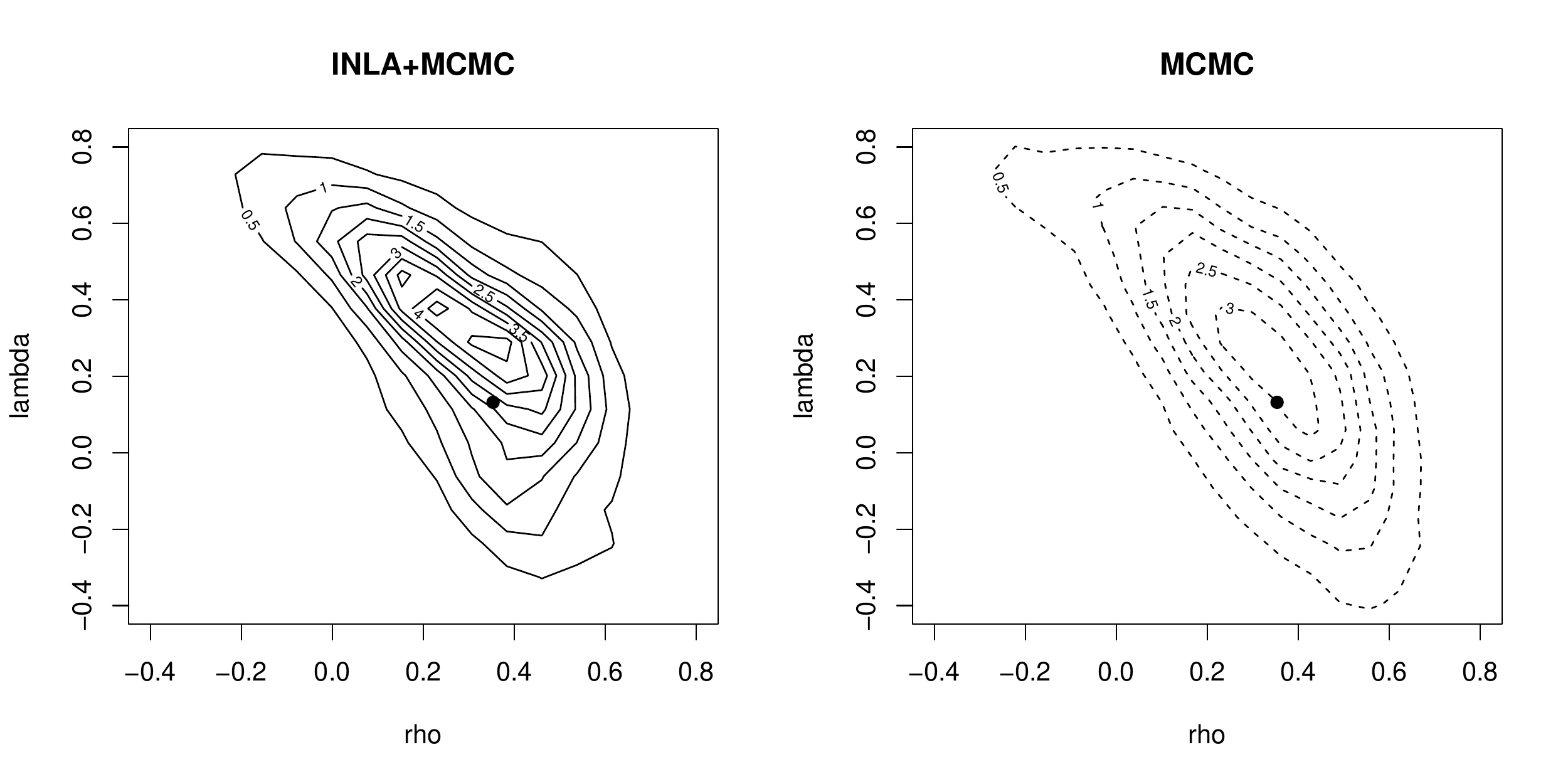}
\caption{Joint posterior distribution of $(\rho, \lambda)$. The solid
dot represents the maximum likelihood estimate.}
\label{fig:contour}
\end{figure}

Finally, the joint posterior distribution of $(\rho, \lambda)$ is shown in
Figure~\ref{fig:contour}. The bivariate distributions obtained with INLA within
MCMC and MCMC look alike. The maximum likelihood estimate has also been added
to the plots (as a black dot), and in both plots looks close to the mode of the
joint posterior distribution. 

\subsubsection{Computation of the impacts}

Spill over effects, or how changes in the covariates in one area
affect its neighbors, are of interest in spatial econometrics. 
These measures are often called \textit{impacts} and they are defined
using partial derivatives as

\begin{equation}
\frac{\partial y_i}{\partial x_{jr}}\ i,j=1\ldots,n;\ r = 1, \ldots, p
\end{equation}
\noindent
Each covariate will have an associated matrix of impacts which, for the
model that we are considering, is:

\begin{equation}
S_r(W) = [I - \rho W]^{-1} (\beta_r + W \gamma_r),\ r = 1,\ldots, p
\end{equation}
\noindent
Note that each element in that matrix will measure how a change in covariate
$r$ in area $j$ will impact on the response in area $i$. These are often
summarized as the average \textit{direct}, \textit{indirect} and \textit{total}
impacts.

Direct impacts measure changes in the response in the same area where the
change occurs, indirect impacts measure changes in adjacent areas and total
impacts are the sum of direct and indirect impacts.  For this reason the
average direct impact is defined as the mean of the trace of the impacts
matrix, the average indirect impact is the sum of all the off-diagonal impacts
divided by $n$ and the total impact is the sum of all the elements in the
impacts matrix divided by $n$.

Following the details in \citet{LeSagePace:2009} and
\citet{GomezRubioetal:2016}, the average total impact for the Manski model for
a covariate $r$ is

\begin{equation} \frac{1}{1 - \rho}\beta_r, \end{equation}
\noindent
and the average direct impact is 

\begin{equation}
n^{-1} tr((I - \rho W)^{-1})\beta_r + n^{-1} tr((I - \rho W)^{-1}W)\gamma_r.
\end{equation}
\noindent
Here, $tr(\cdot)$ denotes the trace of a matrix.  Average indirect impacts can
be obtained by taking the difference between total and direct impacts.

As discussed in \citet{GomezRubioetal:2016}, computing the impacts requires
multivariate inference as the distribution depends on the (joint) posterior
distribution of $\rho$, $\beta_r$ and $\gamma_r$. Hence, computing the impacts
from the posterior marginals of these parameters alone is not enough to obtain
accurate estimates.  Note that in our particular case we are not considering
lagged covariates and that $\gamma_r = 0$, which simplifies the computation of
the impacts.

To compute the impacts we can exploit the fact that we have different models
conditioned on the values of $\rho$. Hence, we could compute the marginal
distribution of the impacts for each value of $\rho$ obtained with the
Metropolis-Hastings and conditional (on $\rho$) marginal $\pi(\beta_r|y,
\rho)$. Then we could average over all the conditional marginal distributions
of the impacts given $rho$ to obtain the final distribution of the impacts,
which could be used to compute summary statistics.

Table~\ref{tab:impacts} summarizes the point estimates and standard deviations
of the different average impacts considered in this example. The estimation
methods for which the impacts are reported are MCMC, INLA within MCMC and
maximum likelihood.  MCMC and INLA within MCMC provided very similar results,
although the latter seems to provide estimates with slightly smaller standard
deviations.

\begin{table}
\begin{small}
\begin{tabular}{l|ccc|ccc}
 & \multicolumn{3}{c|}{INC} & \multicolumn{3}{c}{HOVAL}\\
\hline
Method & Direct & Indirect & Total & Direct & Indirect & Total\\
\hline
MCMC &  -0.96 (0.37) & -0.46 (0.42) & -1.42 (0.66) & 
  -0.30 (0.10) & -0.14 (0.14) & - 0.44 (0.20) \\ 
INLA+MCMC & -0.97 (0.33) & -0.44 (0.35) & -1.40 (0.48) &
  -0.30 (0.10) & -0.13 (0.10) & -0.43 (0.14)\\
Max. lik. & -1.10 & -0.55 & -1.65 & -0.29 & -0.15 & -0.44\\
\end{tabular}

\caption{Mean and standard deviation (in parenthesis) of the impacts
for the Manski model without lagged covariates fitted to the  Columbus dataset.
Covariates are housing value (HOVAL) and household income (INC).}
\label{tab:impacts}
\end{small}
\end{table}

\subsection{Joint Modeling of Three Diseases}
\label{subsec:dismap}

Spatial models have been used for a long time in disease mapping to study
the spatial variation of disease
\citep{BlangiardoCameletti:2015}. Several authors have used INLA to fit spatial
and spatio-temporal models for disease mapping of a single disease as MCMC
methods are often very slow when the number of areas is large.  Modeling
several diseases is more complex and it often involves non-linear terms on some
parameters in the linear predictor of the model
\citep{Lerouxetal:1999,Bivandetal:2015}.  \citet{HeldBest:2001} have considered
the joint modeling of two diseases by including a shared spatial
effect (with different weights for each disease) plus specific spatial terms for
each disease.

In the next example we extend this model to three diseases following
\citet{Downingetal:2008}. They fit a joint model to six diseases with different
weighted shared and specific spatial components. In our case, we will only
consider three diseases, which are modeled using a shared spatial term and
specific spatial patterns.

In particular, let us assume that we have a study region divided into $n$ 
smaller areas and that we are interested in modeling $D$ diseases.
Observed counts of disease $d$ in area $i$ will be denoted by $O^{(d)}_i$.
Similarly, $E^{(d)}_i$ will represent the expected cases of disease $d$
in area $i$.
The model that we will be fitting is 

\begin{equation}
O^{(d)}_i \sim Po(E^{(d)}_i \theta^{(d)}_{i})
\end{equation}
\noindent
Here, $\theta^{(d)}_{i}$ is the relative risk for disease $d$
in area $i$, which is modeled as

\begin{eqnarray}
\log(\theta^{(d)}_{i}) = \alpha^{(d)} + v_i \cdot \delta^{(d)} +s^{(d)}_i;\ 
i=1,\ldots,n;\ d=1,\ldots, D
\end{eqnarray}
\noindent
Parameter $\alpha^{(d)}$ is a disease-specific intercept and it sets the
baseline of the log-relative risks for disease $d$, $v_i$ is a shared component
to all diseases, $\delta^{(d)}$ is a parameter that controls how the shared
component affects disease $d$ and $s^{(d)}_i$ are specific components to each
disease.  The shared and specific components can include different types of
fixed and random effects. \citet{HeldBest:2001} use a partition model to split
areas into regions of similar risk, but other effects can be considered
\citep[see, for example,][]{geobugs:2004}.

Note that this model looks like a standard log-linear model and that, for a
fixed value of $\delta^{(d)}$, it could be fitted with some standard software
packages for Bayesian inference, including \pkg{R-INLA}.  We will use our new
approach to fit this model by taking $\theta_c =
(\delta)^{(1)},\ldots,\delta^{(D)})$.  In \pkg{R-INLA}, this value can be
included as a weight of the effects in the linear predictor using a latent
effect of type \code{besag}.

The prior for $\delta^{(d)}$ will be a log-Normal distribution with mean 0 and
a precision equal to 0.1 to provide vague prior information.  Shared component
$v_i$ will be an intrinsic CAR spatial effect with precision $\tau_v$. Specific
components will also have an intrinsic CAR spatial effect with common precision
$\tau_s$ (which can be easily modeled using the \code{replicate} feature in
\pkg{R-INLA}).

%FIXME: Add priors
%** PRIORS **

We will fit this model to the cases of lip, oral cavity and pharynx tumors
(ICD-10 C00-C14),  esophagus tumor (ICD-10 C15) and stomach tumor (ICD-10
C16) from 1996 to 2014 at the province level in mainland Spain. Expected cases
were computed using age-sex standardized rates from period 1996 - 2014
(excluding 1997).  Population in 1997 was not available but expected cases for
this year were interpolated using the temporal series of expected cases.
Then all expected cases were rescaled so that their sum was equal
to the total number of cases in the whole period.

\begin{figure}[h]
\centering
\includegraphics[width=11cm]{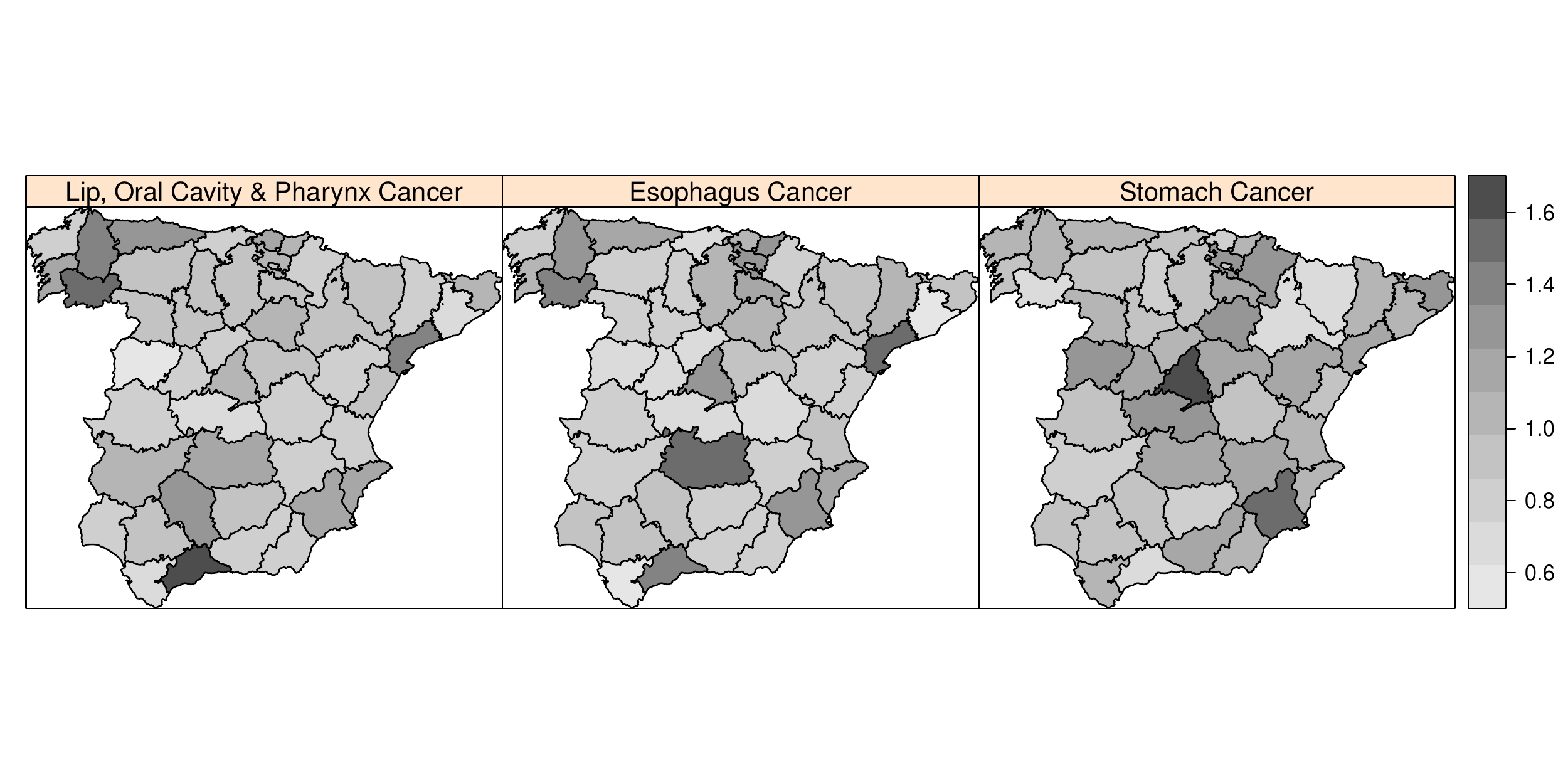}
\caption{Standardized Mortality Ratio of lip, oral cavity and pharynx cancers, 
esophagus cancer and stomach cancer in Spain in 2013.}
\label{fig:SMR}
\end{figure}

Standardized Mortality Ratios (i.e., $O^{(d)}_i / E^{(d)}_i$) for each disease
are shown in the maps in Figure \ref{fig:SMR}. Stomach tumors seems to have a
spatial pattern that is different, and the other two diseases seem to have a
very similar spatial pattern.  We expect that the model described earlier would
capture this joint spatial pattern and account for the different overall rates
by means of the disease specific intercepts included in the specific
components.

\begin{figure}[h]
\centering
\includegraphics[width=11cm]{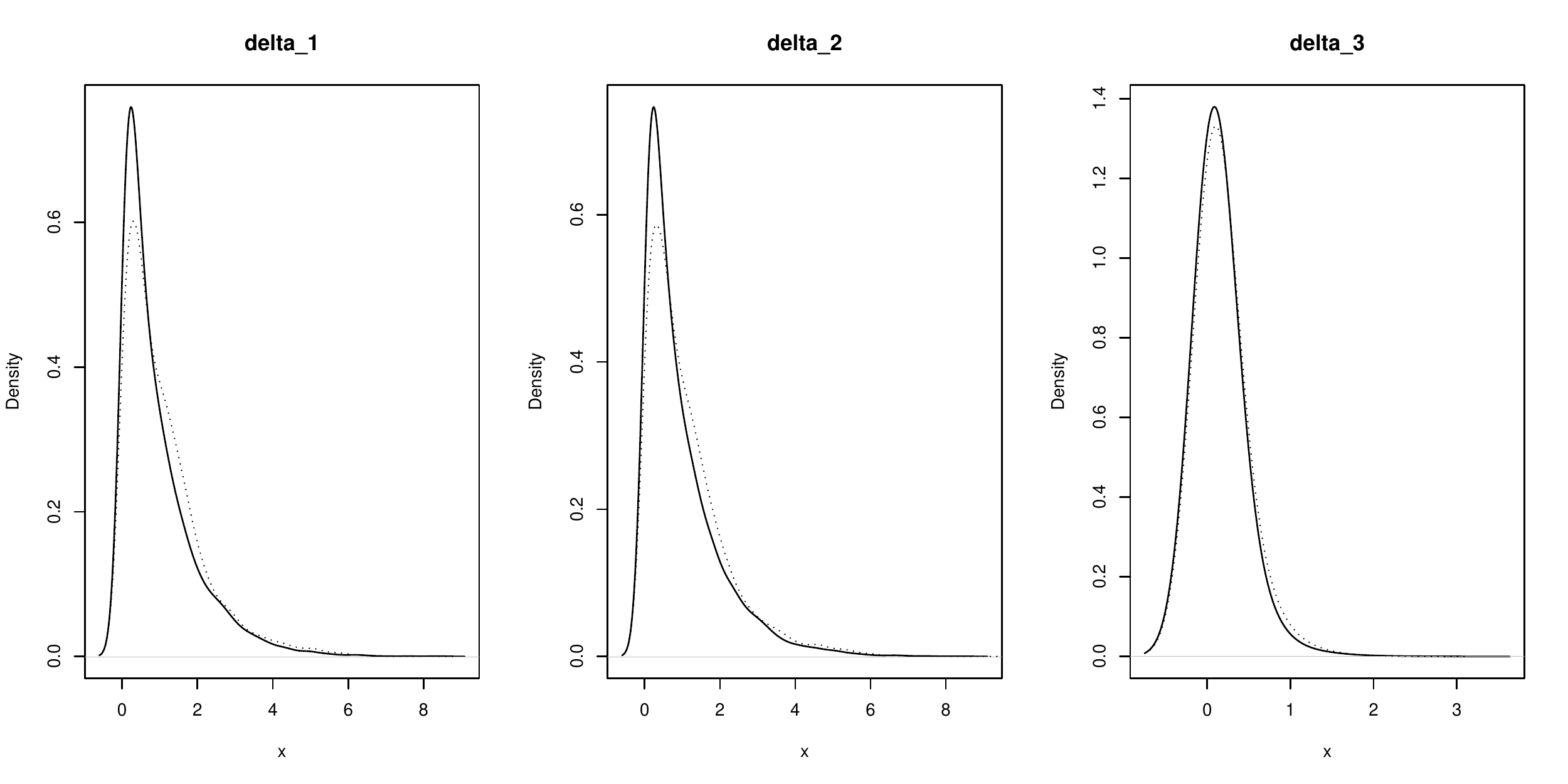}
\centering
\includegraphics[width=10cm]{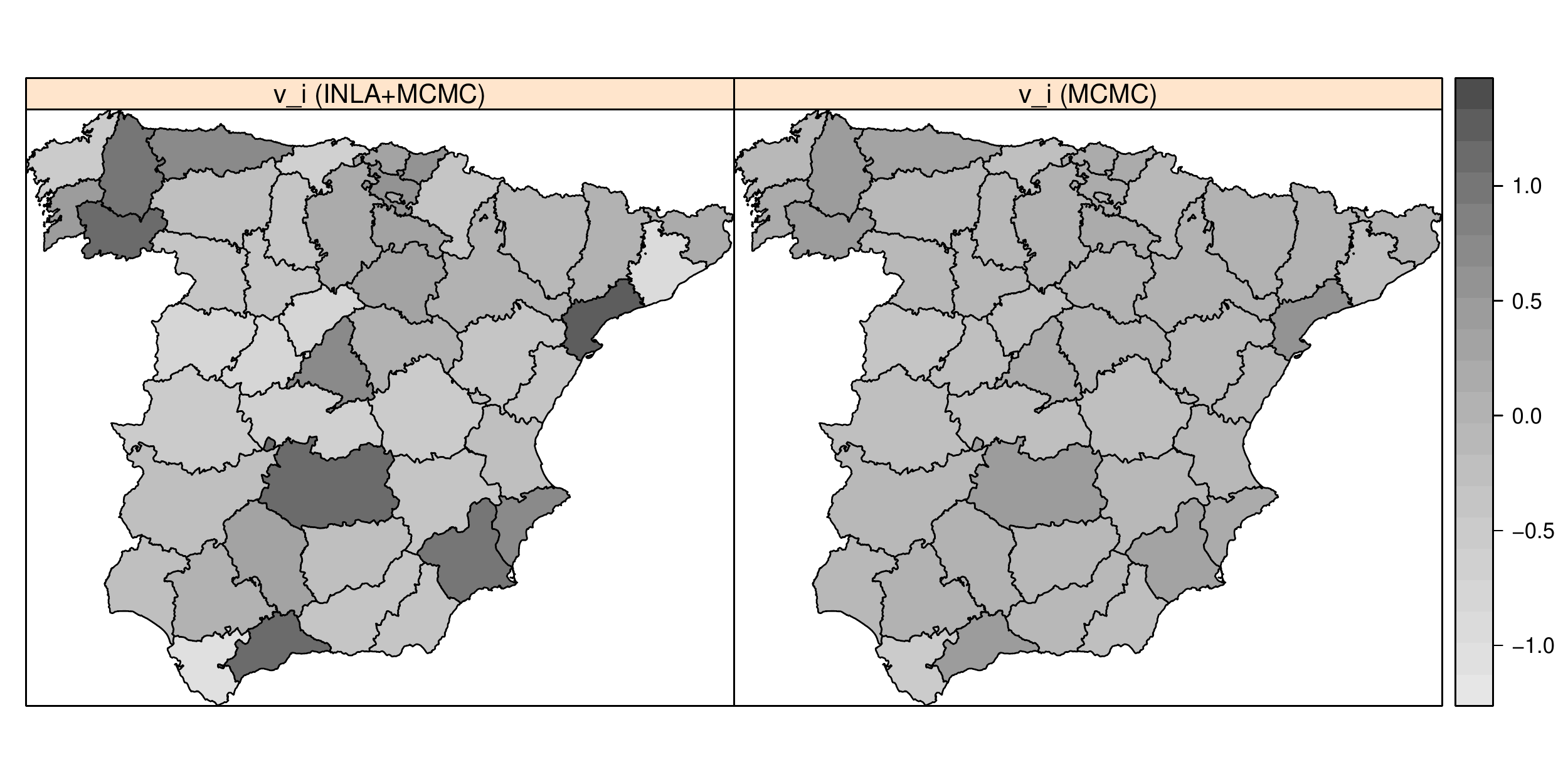}
\caption{Summary of the posterior marginals of $\delta^{(d)}$ (top line)
and the spatial common effect $v_i$ (bottom line). Solid lines represent
estimates obtained with INLA within MCMC and the dashed line represents
the marginals obtained with MCMC.}
\label{fig:shared}
\end{figure}

Figure \ref{fig:shared} summarizes the estimates obtained with MCMC (Gibbs
sampling with WinBUGS) and INLA within MCMC. In all plots, the solid lines
represent the estimates of the posterior distribution using INLA within MCMC
whilst the dashed lines is the estimate using MCMC.  The top row shows the
estimates of the posterior distribution of weights $\delta^{(d)}$ for the three
diseases for which there is a very good agreement between both estimation
methods. In the bottom line we can find  the estimates of the common spatial
effect, for which there is also good agreement between INLA within MCMC and
MCMC.

\begin{figure}[h]
\centering
\includegraphics[width=11cm]{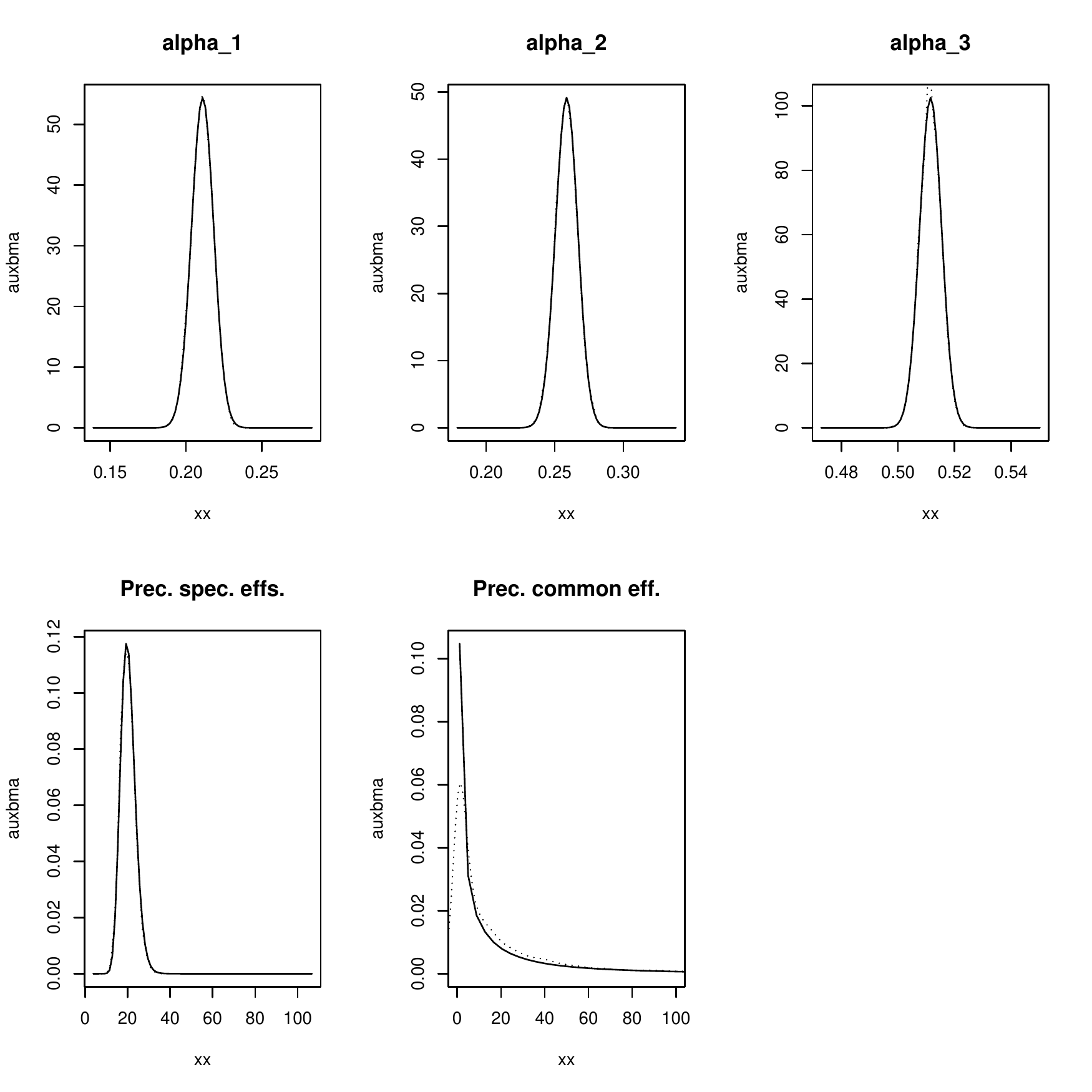}
\caption{Marginal distributions of the parameters in the model
obtained by Bayesian model averaging the conditional marginals
obtained at each step of the Metropolis-Hastings algorithm.}
\label{fig:params}
\end{figure}

Finally, Figure~\ref{fig:params} displays the marginal distributions of the
disease specific intercepts $\alpha^{(d)}$ and precisions that appear in the
model. These have been obtained by doing an average of the conditional
marginals obtained at each step of the Metropolis-Hastings algorithm. In
general, there is good agreement between our approach and MCMC.

Although it is not shown here, we have observed a positive strong correlation
between $\delta^{(1)}$ and $\delta^{(2)}$ We believe that this happens because
of the similar spatial pattern that both diseases have.  $\delta^{(3)}$ shows
less correlation with the other two weighing parameters.  This kind of
multivariate inference has also been possible because we have been able to
estimate the joint posterior distribution of $(\delta^{(1)}, \delta^{(2)},
\delta^{(3)})$ by fitting the model with INLA within MCMC.

\section{Discussion}
\label{sec:discussion}

\citet{GomezRubioRue:2016} develop a novel approach to fit Bayesian models not
implemented in the \pkg{R-INLA} by combining Metropolis-Hastings and INLA. In
this paper we have exploited this method to show how to fit complex spatial
models with several spatial components so that multivariate inference on a small ensemble
of important parameters is feasible.

As seen in the examples presented in this paper, INLA within Metropolis can be
used to fit complex spatial models with a simple implementation of the
Metropolis-Hastings algorithm. This implementation is simpler compared to a
full implementation of the model using several MCMC algorithms as sampling only
involves a few parameters. Furthermore, convergence of the MCMC simulations
is achieved in less iterations because the number of parameters
that need to be simulated from is greatly reduced compared to fitting the model
completely using MCMC.  For large datasets, this should be an interesting
alternative as INLA can fit conditional models faster than MCMC.

For cases in which the spatial model can be fitted completely with
\pkg{R-INLA}, the approach presented in this paper is also appealing because it
makes multivariate inference based on the joint posterior possible.

\section*{Acknowledgments}

This work has been supported by grant PPIC-2014-001, funded by Consejer\'ia de
Educaci\'on, Cultura y Deportes (JCCM) and FEDER, and grant MTM2016-77501-P,
funded by Ministerio de Econom\'ia y Competitividad.

\section*{Bibliography}

\bibliographystyle{chicago}
\bibliography{local}

\begin{thebibliography}{}

\bibitem[\protect\citeauthoryear{Anselin}{Anselin}{1988}]{Anselin:1988}
Anselin, L. (1988).
\newblock {\em Spatial {E}conometrics: Methods and Models}.
\newblock Dordrecht: Kluwer.

\bibitem[\protect\citeauthoryear{Besag, York, and Mollie}{Besag
  et~al.}{1991}]{Besagetal:1991}
Besag, J., J.~York, and A.~Mollie (1991).
\newblock Bayesian image restoration, with two applications in spatial
  statistics.
\newblock {\em Annals of the Institute of Statistical Mathematics\/}~{\em
  43\/}(1), 1--59.

\bibitem[\protect\citeauthoryear{Bivand and Piras}{Bivand and
  Piras}{2015}]{BivandPiras:2015}
Bivand, R. and G.~Piras (2015).
\newblock Comparing implementations of estimation methods for spatial
  econometrics.
\newblock {\em Journal of Statistical Software\/}~{\em 63\/}(1), 1--36.

\bibitem[\protect\citeauthoryear{Bivand, G\'omez-Rubio, and Rue}{Bivand
  et~al.}{2014}]{Bivandetal:2014}
Bivand, R.~S., V.~G\'omez-Rubio, and H.~Rue (2014).
\newblock Approximate {B}ayesian inference for spatial econometrics models.
\newblock {\em Spatial Statistics\/}~{\em 9}, 146--165.

\bibitem[\protect\citeauthoryear{Bivand, G\'omez-Rubio, and Rue}{Bivand
  et~al.}{2015}]{Bivandetal:2015}
Bivand, R.~S., V.~G\'omez-Rubio, and H.~Rue (2015).
\newblock Spatial data analysis with \proglang{R}-\pkg{INLA} with some
  extensions.
\newblock {\em Journal of Statistical Software\/}~{\em 63\/}(20), 1--31.

\bibitem[\protect\citeauthoryear{Blangiardo and Cameletti}{Blangiardo and
  Cameletti}{2015}]{BlangiardoCameletti:2015}
Blangiardo, M. and M.~Cameletti (2015).
\newblock {\em Spatial and Spatio-temporal Bayesian Models with R - INLA}.
\newblock John Wiley \& Sons, Inc.

\bibitem[\protect\citeauthoryear{Blangiardo, Cameletti, Baio, and
  Rue}{Blangiardo et~al.}{2013}]{Blangiardoetal:2013}
Blangiardo, M., M.~Cameletti, G.~Baio, and H.~Rue (2013).
\newblock Spatial and spatio-temporal models with \pkg{R-INLA}.
\newblock {\em Spatial and Spatio-Temporal Epidemiology\/}~{\em 4}, 33 -- 49.

\bibitem[\protect\citeauthoryear{Brooks, Gelman, Jones, and Meng}{Brooks
  et~al.}{2011}]{MCMC:2011}
Brooks, S., A.~Gelman, G.~Jones, and X.~Meng (Eds.) (2011).
\newblock {\em Handbook of Markov Chain Monte Carlo}.
\newblock Chapman \& Hall/CRC Handbooks of Modern Statistical Methods. CRC
  Press.

\bibitem[\protect\citeauthoryear{Cameletti, G\'omez-Rubio, and
  Blangiardo}{Cameletti et~al.}{2016}]{Camelettietal:2016}
Cameletti, M., V.~G\'omez-Rubio, and M.~Blangiardo (2016).
\newblock Missing data analysis with the integrated nested laplace
  approximation.
\newblock {\em In preparation\/}.

\bibitem[\protect\citeauthoryear{Downing, Forman, Gilthorpe, Edwards, and
  Manda}{Downing et~al.}{2008}]{Downingetal:2008}
Downing, A., D.~Forman, M.~S. Gilthorpe, K.~L. Edwards, and S.~O. Manda (2008).
\newblock Joint disease mapping using six cancers in the yorkshire region of
  england.
\newblock {\em International Journal of Health Geographics\/}~{\em 7\/}(41),
  1--14.

\bibitem[\protect\citeauthoryear{Gilks, Gilks, Richardson, and
  Spiegelhalter}{Gilks et~al.}{1996}]{Gilksetal:1996}
Gilks, W., W.~Gilks, S.~Richardson, and D.~Spiegelhalter (1996).
\newblock {\em Markov Chain Monte Carlo in practice}.
\newblock Boca Raton, Florida: Chapman \& Hall.

\bibitem[\protect\citeauthoryear{G\'omez-Rubio, Bivand, and Rue}{G\'omez-Rubio
  et~al.}{2014}]{gomez-rubioetal14}
G\'omez-Rubio, V., R.~S. Bivand, and H.~Rue (2014).
\newblock Spatial models using laplace approximation methods.
\newblock In M.~M. Fischer and P.~Nijkamp (Eds.), {\em Handbook of Regional
  Science}, pp.\  1401--1417. Berlin: Springer.

\bibitem[\protect\citeauthoryear{G\'omez-Rubio, Bivand, and Rue}{G\'omez-Rubio
  et~al.}{2016}]{GomezRubioetal-slm:2016}
G\'omez-Rubio, V., R.~S. Bivand, and H.~Rue (2016).
\newblock Estimating spatial econometrics models with integrated nested laplace
  approximation.
\newblock {\em In preparation\/}.

\bibitem[\protect\citeauthoryear{G\'omez-Rubio, Cameletti, and
  Finazzi}{G\'omez-Rubio et~al.}{2015}]{GomezRubioetal:2015}
G\'omez-Rubio, V., M.~Cameletti, and F.~Finazzi (2015).
\newblock Analysis of massive marked point patterns with stochastic partial
  differential equations.
\newblock {\em Spatial Statistics\/}~{\em B\/}(14), 176--196.

\bibitem[\protect\citeauthoryear{G\'omez-Rubio and Rue}{G\'omez-Rubio and
  Rue}{2016}]{GomezRubioRue:2016}
G\'omez-Rubio, V. and H.~Rue (2016).
\newblock Markov chain monte carlo with the integrated nested laplace
  approximation.
\newblock {\em In preparation\/}.

\bibitem[\protect\citeauthoryear{Gómez-Rubio, Bivand, and Rue}{Gómez-Rubio
  et~al.}{2016}]{GomezRubioetal:2016}
Gómez-Rubio, V., R.~S. Bivand, and H.~Rue (2016).
\newblock Estimating spatial econometrics models with integrated nested laplace
  approximation.
\newblock {\em In preparation\/}.

\bibitem[\protect\citeauthoryear{Haining}{Haining}{2003}]{Haining:2003}
Haining, R. (2003).
\newblock {\em {Spatial Data Analysis: Theory and Practice}}.
\newblock Cambridge University Press.

\bibitem[\protect\citeauthoryear{Hastings}{Hastings}{1970}]{Hastings:1970}
Hastings, W.~K. (1970).
\newblock {Monte Carlo} sampling methods using {Markov} chains and their
  applications.
\newblock {\em Biometrika\/}~{\em 57}, 97 -- 109.

\bibitem[\protect\citeauthoryear{Knorr-Held and Best}{Knorr-Held and
  Best}{2001}]{HeldBest:2001}
Knorr-Held, L. and N.~Best (2001).
\newblock A shared component model for detecting joint and selective clustering
  of two diseases.
\newblock {\em Journal of the Royal Statistical Society, Series A\/}~{\em
  1\/}(164), 73 -- 85.

\bibitem[\protect\citeauthoryear{Leroux, Lei, and Breslow}{Leroux
  et~al.}{1999}]{Lerouxetal:1999}
Leroux, B., X.~Lei, and N.~Breslow (1999).
\newblock Estimation of disease rates in small areas: A new mixed model for
  spatial dependence.
\newblock In M.~Halloran and D.~Berry (Eds.), {\em Statistical Models in
  Epidemiology, the Environment and Clinical Trials}, pp.\  135--178. New York:
  Springer-Verlag.

\bibitem[\protect\citeauthoryear{{LeSage} and Pace}{{LeSage} and
  Pace}{2009}]{LeSagePace:2009}
{LeSage}, J. and R.~K. Pace (2009).
\newblock {\em Introduction to Spatial Econometrics}.
\newblock Chapman and Hall/CRC.

\bibitem[\protect\citeauthoryear{Li, Brown, Rue, {al-Maini}, and Fortin}{Li
  et~al.}{2012}]{Lietal:2012}
Li, Y., P.~Brown, H.~Rue, M.~{al-Maini}, and P.~Fortin (2012).
\newblock Spatial modelling of {Lupus} incidence over 40 years with changes in
  census areas.
\newblock {\em Journal of the Royal Statistical Society, Series C\/}~{\em 61},
  99--115.

\bibitem[\protect\citeauthoryear{Lindgren and Rue}{Lindgren and
  Rue}{2015}]{LindgrenRue:2015}
Lindgren, F. and H.~Rue (2015).
\newblock Bayesian spatial modelling with r-inla.
\newblock {\em Journal of Statistical Software\/}~{\em 63\/}(1), 1--25.

\bibitem[\protect\citeauthoryear{Lindgren, Rue, and Lindstrom}{Lindgren
  et~al.}{2011}]{Lindgrenetal:2011}
Lindgren, F., H.~Rue, and J.~Lindstrom ({2011}).
\newblock {An explicit link between Gaussian fields and Gaussian Markov random
  fields: the stochastic partial differential equation approach}.
\newblock {\em Journal of the Royal Statistical Society, Series B\/}~{\em
  {73}\/}({Part 4}), {423--498}.

\bibitem[\protect\citeauthoryear{Manski}{Manski}{1993}]{Manski:1993}
Manski, C.~F. (1993).
\newblock Identification of endogenous social effects: the reflection problem.
\newblock {\em Review of Economic Studies\/}~{\em 60\/}(3), 531--542.

\bibitem[\protect\citeauthoryear{Martins, Simpson, Lindgren, and Rue}{Martins
  et~al.}{2013}]{Martinsetal:2013}
Martins, T.~G., D.~Simpson, F.~Lindgren, and H.~Rue (2013).
\newblock {Bayesian computing with INLA: New features}.
\newblock {\em Computational Statistics \& Data Analysis\/}~{\em 67}, 68--83.

\bibitem[\protect\citeauthoryear{Metropolis, Rosenbluth, Rosenbluth, Teller,
  and Teller}{Metropolis et~al.}{1953}]{Metropolisetal:1953}
Metropolis, N., A.~W. Rosenbluth, M.~N. Rosenbluth, A.~H. Teller, and E.~Teller
  (1953).
\newblock Equations of state calculations by fast computing machine.
\newblock {\em Journal of Chemical Physics\/}~{\em 21}, 1087 -- 1091.

\bibitem[\protect\citeauthoryear{Plummer}{Plummer}{2016}]{rjags:2016}
Plummer, M. (2016).
\newblock {\em rjags: Bayesian Graphical Models using MCMC}.
\newblock R package version 4-6.

\bibitem[\protect\citeauthoryear{{R Core Team}}{{R Core Team}}{2016}]{R:2016}
{R Core Team} (2016).
\newblock {\em R: A Language and Environment for Statistical Computing}.
\newblock Vienna, Austria: R Foundation for Statistical Computing.

\bibitem[\protect\citeauthoryear{Rue and Held}{Rue and
  Held}{2005}]{RueHeld:2005}
Rue, H. and L.~Held (2005).
\newblock {\em Gaussian Markov Random Fields. Theory and Applications}.
\newblock Chapman \& Hall/CRC.

\bibitem[\protect\citeauthoryear{Rue, Martino, and Chopin}{Rue
  et~al.}{2009}]{isi:000264374200002}
Rue, H., S.~Martino, and N.~Chopin ({2009}).
\newblock {Approximate Bayesian inference for latent Gaussian models by using
  integrated nested Laplace approximations}.
\newblock {\em Journal of the Royal Statistical Society, Series B\/}~{\em
  {71}\/}({Part 2}), {319--392}.

\bibitem[\protect\citeauthoryear{Simpson, Illian, Lindgren, Sørbye, and
  Rue}{Simpson et~al.}{2016}]{Simpsonetal:2016}
Simpson, D., J.~B. Illian, F.~Lindgren, S.~H. Sørbye, and H.~Rue (2016).
\newblock Going off grid: computationally efficient inference for log-gaussian
  cox processes.
\newblock {\em Biometrika\/}~(To appear).

\bibitem[\protect\citeauthoryear{Thomas, Best, Lunn, Arnold, and
  Spiegelhalter}{Thomas et~al.}{2004}]{geobugs:2004}
Thomas, A., N.~Best, D.~Lunn, R.~Arnold, and D.~Spiegelhalter (2004).
\newblock {GeoBUGS} user manual.

\bibitem[\protect\citeauthoryear{Ugarte, Adin, Goicoa, and Militino}{Ugarte
  et~al.}{2014}]{Ugarteetal:2014}
Ugarte, M.~D., A.~Adin, T.~Goicoa, and A.~F. Militino (2014).
\newblock On fitting spatio-temporal disease mapping models using approximate
  bayesian inference.
\newblock {\em Statistical Methods in Medical Research\/}~{\em 23\/}(6),
  507--530.

\end{thebibliography}
%INLA_econometrics,RS_handbook_SpatStat_INLA,INLAMCMC,extra}

\end{document}